\documentclass[twocolumn,showpacs,aps]{revtex4}

\usepackage{epsfig}

\begin{document}
\author{M. Shuker\footnotemark \footnotetext{Email
address: shuker@physics.technion.ac.il}, A. Ben-kish, R. A.
Nemirovsky, A. Fisher and  A. Ron} \affiliation{Department of
Physics, Technion - Israel Inst. of Tech., Haifa 32000, Israel}
\title{The effects of the pre-pulse on capillary discharge extreme ultraviolet laser}
\begin{abstract}
In the past few years collisionally pumped extreme ultraviolet
(XUV) lasers utilizing a capillary discharge were demonstrated. An
intense current pulse is applied to a gas filled capillary,
inducing magnetic collapse (\emph{Z-pinch}) and formation of a
highly ionized plasma column. Usually, a small current pulse
(pre-pulse) is applied to the gas in order to pre-ionize it prior
to the onset of the main current pulse. In this paper we
investigate the effects of the pre-pulse on a capillary discharge
Ne-like Ar XUV laser ($46.9nm$). The importance of the pre-pulse
in achieving suitable initial conditions of the gas column and
preventing instabilities during the collapse is demonstrated.
Furthermore, measurements of the amplified spontaneous emission
(ASE) properties (intensity, duration) in different pre-pulse
currents revealed unexpected sensitivity. Increasing the pre-pulse
current by a factor of two caused the ASE intensity to decrease by
an order of magnitude - and to nearly disappear. This effect is
accompanied by a slight increase in the lasing duration. We
attribute this effect to axial flow in the gas during the
pre-pulse.
\end{abstract}

\maketitle
\begin{figure}
    \epsfig{file=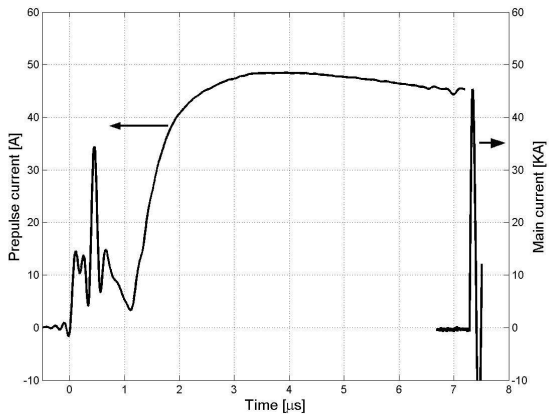}
    \caption{The pre-pulse current and the main current. The first peak in the pre-pulse current is the spark followed by current from a capacitor discharge. In this case the peak pre-pulse current is ~50A. The main current pulse starts ~6$\mu$s after the pre-pulse.}
    \label{FigurePrepulseCurrent}
\end{figure}

\begin{figure}
    \epsfig{file=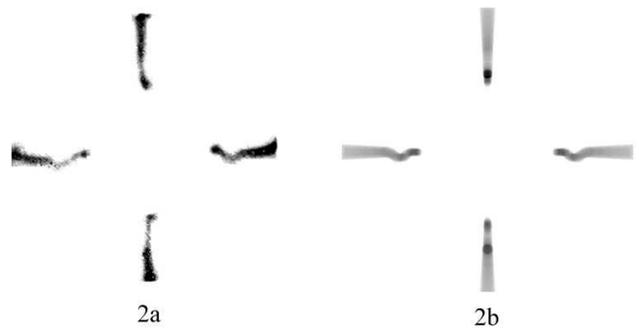}
    \caption{Effect of low current pre-pulse on the stability of the plasma collapse during the discharge. a) An off-axis pinhole image of the collapsing gas column with 5A peak current (taken from four different directions). b) A reconstruction of the instability of the collapse using a simple ray-tracing model.}
    \label{FigurePhkinkComparison}
\end{figure}

 In the last decades the possibility of achieving
Amplified Spontaneous Emission (ASE) in the soft X-ray and XUV
regimes was extensively explored \cite{Elton1990}. One of the
realizations utilizes a homogenous column of highly ionized plasma
created by a fast capillary discharge \cite{Rocca1994}.
Specifically, a fast ($\sim50ns$) and intense ($\sim50kA$) current
pulse is applied to a capillary filled with low pressure
($\sim1Torr$) argon gas. The current induces magnetic forces that
attract the gas towards the capillary axis. During the
self-collision of the gas on the axis a column of highly ionized
Ar plasma is formed (this technique to create high temperature
plasmas is called Z-Pinch). Careful design of the experimental
parameters results in high abundance of Ne-like Ar ions in the
plasma, and population inversion between the $3P$ and $3S$
electronic configurations. The population inversion is due to
excitations by collisions with hot electrons in the plasma and
spontaneous emission - i.e. the \textsl{collisional excitation}
scheme \cite{Elton1990}. Strong amplification was observed in the
$3P\rightarrow 3S$ transition at $46.9nm$
\cite{Rocca1994}-\cite{Tomassetti2002}. The timing of the
amplification (that lasts about 1ns) is close to the pinch time
(i.e. the collision of shock waves on the capillary axis).
Obviously, for efficient amplification the plasma column must be
relatively stable and homogenous (at least up to the amplification
time). Several effects might cause instabilities or inhomogeneity
in the Z-pinch process (see Ref. \cite{Ryutov2000} and references
therein). One of the possible reasons for inhomogeneity is the
initial electrical breakdown through the gas column. At the onset
of the main current pulse, high voltage across the capillary
(hundreds of kV), may result in channel-like breakdown along the
capillary walls (channel sparks) \cite{Rees1973}. A technique to
overcome this inhomogeneity is to pre-ionize the gas by a slow,
low intensity current pulse ("pre-pulse"). In Refs.
\cite{Rocca1994}-\cite{Tomassetti2002} a small current (typically
$10-100A$) is applied for a few $\mu s$ before the onset of the
main current pulse. The pre-pulse current creates sufficient
amount of pre-ionization in the argon gas, allowing the main
current pulse to flow homogenously (i.e. with an axial symmetry).
In this paper we report experimental results showing the effect of
the pre-pulse current on the operation of capillary discharge
X-ray lasers. We investigate the homogeneity of the gas column
during the pre-pulse itself, and during the plasma collapse.
Finally we show the influence of the pre-pulse on the laser pulse
properties.\\
Our experimental setup consists of a $1MV$ Marx type generator
connected to a $7\Omega$ transmission line. A self triggered spark
gap connects the line and the capillary load ($5mm$ inner
diameter, filled with $500mTorr$ of Ar gas). A sine-shaped, $50KA$
peak current pulse is typically used with $100-120ns$ half cycle
time (depending on the length of the capillary). Before the onset
of the main current pulse a so-called "pre-pulse" current is
applied. The pre-pulse current is created by a discharge of a
small capacitor and has a typical RC shape with a decay time of
$\sim 20\mu s$ - see figure \ref{FigurePrepulseCurrent}.
The maximum current of the pre-pulse is controlled by a current
limiting resistor. We note that prior to the pre-pulse a small
spark was initiated at one end of the capillary ("pre-pre-pulse").
This spark, initiated across a $0.3mm$ long surface of alumina,
increases the number of electrons in the gas providing easy
breakdown conditions for the pre-pulse. In the experiments
reported in this paper we used glass or alumina capillaries with
various length ($5-16.5cm$). The exact properties of the capillary
are stated before the description of each experiment.\\
Several measurements were carried out to investigate the role of
pre-pulse current in the dynamics of the discharge and lasing
processes. First, we measured the uniformity of the gas column
illumination during the pre-pulse by means of visible light fast
photography \cite{BenKish1998}. A $5cm$ long glass capillary was
used in these experiments to allow side-view imaging. The visible
light emission of the gas in the capillary is indicative of
current flow, ionization and heating of the gas. Therefore a
non-uniform emission indicates that the volume of the gas is
heated in a non-uniform manner. Experiments were carried out with
pre-pulse peak currents in the $5-50A$ range, and the emission was
monitored up to $20\mu s$ after the pre-pulse started. A $5A$
pre-pulse current caused a channels-like emission pattern in the
gas (even after $20\mu s$ the emission of the gas was
inhomogeneous). On the contrary, a $50A$ peak current pre-pulse
induced a homogenous emission in the gas. The homogenous
emission was achieved $6\mu s$ after the onset of the pre-pulse current.\\
In order to further investigate the effect of the pre-pulse
current on the stability of the plasma column collapse, an
off-axis pinhole camera was used \cite{BenKish2001}. This camera
consists of an array of four pinholes placed in front of the
capillary, with each of the holes located slightly off the
capillary axis. The position of the pinholes allowed imaging of
the entire length of the plasma column from four different
directions (an $8cm$ long alumina capillary was used in these
experiments). The images from all four pinholes were obtained on a
gated multi-channel plate (with $3ns$ gating time) that was
coupled to a CCD camera. A thin Mylar filter was used to limit the
spectral response of the detector ($\sim2-5nm$), so only radiation
from highly ionized Ar ions ($Ar^{8+},Ar^{9+}$) was measured. We
have performed a series of pinhole measurements with different
pre-pulse conditions. Usually, the image was taken at the instance
the plasma column self-collided on the capillary axis (the pinch
time). The pinhole images were analyzed using a simple ray-tracing
code. Figure \ref{FigurePhkinkComparison}a shows a pinhole image
taken with a relatively low $5A$ peak current pre-pulse,
displaying an instability of the plasma column (the cone shape of
the column is a result of the measurement geometry, causing
different optical magnification along the plasma column
\cite{BenKish1998}).

\begin{figure}
    \epsfig{file=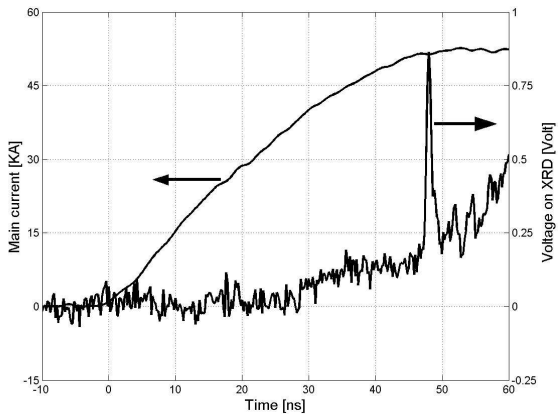}
    \caption{A typical XRD measurement. The graph shows a main current signal and a corresponding XRD signal (for a relatively weak laser). A sharp laser pulse is evident (at $t\simeq48ns$) on top of a slowly increasing radiation of the heating plasma.}
    \label{FigureXRDTypical}
\end{figure}

In figure \ref{FigurePhkinkComparison}b a reconstruction of the
experimental image is demonstrated using a ray-tracing code. The
best reconstruction was achieved by displacing a $2cm$ long
segment in the middle of the plasma column by $150\mu m$ from the
axis. This instability obviously diminishes the amplification as
the gain region diameter is of the order of $100\mu m$
\cite{BenKish2001}. Note that the magnitude of the instability has
a large shot to shot variation and some shots display relatively
stable collapse even with $5A$ pre-pulse current. This is
consistent with the fact that occasionally we measured strong XUV
amplification even with low pre-pulse currents. Repeating the same
measurement with a pre-pulse peak current of $50A$ showed a stable
collapse in every shot. These measurements indicate that in case
the pre-pulse current is too low, an instability develops during
the plasma collapse. Its magnitude depends on the initial
conditions of the
column (that are slightly different in every shot).\\
The two measurements discussed above indicated that minimal
pre-pulse current must be used in order to insure instabilities
free collapse of the plasma column. Indeed, appropriate pre-pulse
conditions resulted in strong lasing at the $3S\rightarrow 3P$
transition of the Ne-like Ar ion \cite{BenKish2001}. A Carbon
X-ray diode was placed in front of the capillary to measure the
temporal history of XUV radiation during the main current pulse.
The response of the Carbon photo-cathode is in the spectral range
$40-100nm$ ($\sim12-30eV$). Special care was taken to ensure high
bandwidth of the entire measurement system (The overall bandwidth
was $\sim1GHz$). The signal was measured on a $5GS/s$ digitizing
oscilloscope.  In this set of experiments we used a $16.5cm$ long
alumina capillary to obtain a laser pulse strong enough to be
easily measured in this time-resolved technique. A typical signal
from the X-ray diode (XRD) shows a combination of two physical
effects. A relatively slow radiation pulse (typical time
$\sim50ns$) starts $\sim20ns$ after the onset of the main current
and peaks after $60-70ns$. On top of the slow pulse an extremely
fast and intense radiation pulse is observed (typical time of
$\sim1ns$). The latter pulse appears $45-50ns$ after the onset of
the main current. We attribute the slow pulse to the radiation of
the heated plasma in the capillary, while the fast pulse is
attributed to the lasing action in the plasma. Figure
\ref{FigureXRDTypical} displays typical main current and XRD
signals.
We have chosen to display an experiment with a relatively weak
laser signal so the
less intense signal of the plasma radiation is evident. \\
While investigating the properties of the laser a strong
sensitivity to the pre-pulse peak current was observed. A series
of consecutive measurements was performed changing only the peak
pre-pulse current (by changing the current limiting resistor in
the pre-pulse circuit). The rest of the experimental parameters
were kept constant. In all the shots reported below the peak of
the main current was $52\pm1kA$ and the Ar pressure in the
capillary was $500\pm10mTorr$. The properties of the laser pulse
were analyzed in the following manner. The background plasma
radiation in the vicinity of the pulse was subtracted, and the
pulse was fitted to a Gaussian (using a least squares minimization
algorithm). The properties of the laser pulse (intensity, time,
FWHM) were deduced from the Gaussian fit. This technique of
analysis is required to reduce the experimental errors caused by
the limited sampling rate. We have investigated the relation
between the laser pulse properties and the pre-pulse peak current.
The timing of the laser pulse (relative to the onset of the main
current) was not influenced by the change in the pre-pulse peak
current (in all the measurements the lasing time was within a
range of $1ns$). However, both the intensity of the laser pulse
and its duration show a systematic dependence on the pre-pulse
peak current.
\begin{figure}
    \epsfig{file=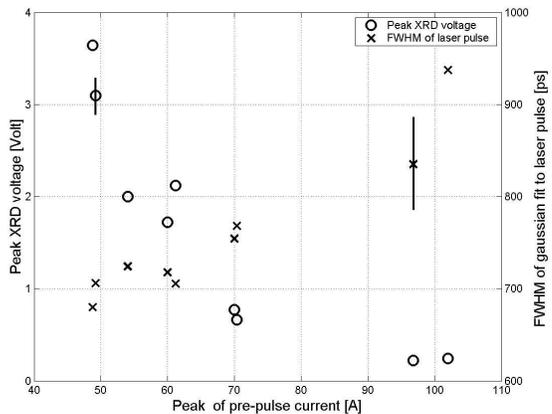}
    \caption{The effect of the pre-pulse peak current on features of the laser pulse. The laser pulse was fitted numerically to a gaussian. The pluses indicate the laser intensity. The squares indicate the duration of the laser pulse.}
    \label{FigureIntenstyDurationVsPrepulse}
\end{figure}
In figure \ref{FigureIntenstyDurationVsPrepulse} the intensity and
duration of the laser pulse are plotted against the pre-pulse peak
current in the range $50-100A$. When the pre-pulse current is
increased - a rapid decrease in the laser pulse intensity is
observed. It is important to note that a relatively small increase
in the pre-pulse (from $50A$ to $100A$) reduced the intensity of
the laser by an order of magnitude. Another experimental
observation is that while the laser intensity
decreases, the duration of the laser pulse increases. \\
The purpose of the pre-pulse current in the capillary discharge
experiment is to heat and ionize the gas column. This
pre-ionization allows the main current to flow uniformly in the
gas column (i.e. with axial symmetry) and prevents channel like
breakdowns. Therefore, it is not surprising that a minimal level
of pre-pulse current is required to prevent non-uniformity
effects. The pre-pulse current, being three orders of magnitude
lower than the main current, is usually assumed to have no effect
on the Z-pinch dynamics (other than insuring a stable collapse).
However we found that increasing the pre-pulse current can reduce
substantially the lasing intensity. The fact that the timing of
the laser pulse stayed constant for all the pre-pulse currents we
tested, suggests that the radial dynamics of the collapse is not
affected by the pre-pulse amplitude (in the range $50-100A$). The
decrease in the intensity of the laser can be attributed to
several effects. In our experimental system the axial magnetic
field is determined by the pre-pulse current, hence the change
might be due to the different axial magnetic field. Another
possibility is that the pre-pulse causes radial perturbations in
the gas column (skin effect, radial motion) that later effects the
main collapse. Finally, the heating of the gas column during the
pre-pulse might cause axial flow of the gas through the hole in
one of the electrodes (through which the laser radiation exits).
The part of the capillary affected by the axial flow will have
different initial gas density, and therefore very low
gain \cite{Rocca1997}. Furthermore, the flow of gas out of the capillary will increase the absorption of the laser radiation when it exits the capillary.\\
The maximal axial magnetic field generated by the pre-pulse
currents tested here is relatively low ($\sim1.5kG$), not in the
range that should effect the lasing \cite{Tomasel1996}. In order
to estimate the radial dynamics during the pre-pulse current a
series of one-dimensional MHD simulations \cite{Nemirovsky1999}
was performed. Even the maximal pre-pulse current used in our
experimental investigation ($100A$) caused negligible radial
non-uniformity in the plasma column (the uniformity was better
than $1.5\%$ during the entire pre-pulse). Finally, the axial
motion during the pre-pulse was estimated by calculating the sound
velocity of the heated gas from the electron temperature found
using the MHD simulation. The simulation results show that during
the $6\mu s,100A$ pre-pulse the argon is heated to
$T_e=T_i\cong1.5eV$ and reach mean ionization of
$\overline{Z}\simeq1$. The ions sound velocity in the plasma is
given by Ref. \cite{Huba2000}:
\begin{equation}
C_i=\sqrt{\frac{\gamma Zk_BT_e}{m_i}}
\end{equation}
Where $\gamma=5/3$ for mono-atomic gases, $Z$ is the ionization
level, $k_B$ is the Boltzman constant, $T_e$ is the electronic
temperature and $m_i$ is the mass of the ion. The calculated sound
velocity of singely ionized ions is therefore
$C_i\cong2500m/s=2.5mm/\mu s$. During our $6\mu s, 100A$ pre-pulse
a length of $1.5cm$ of the capillary is affected by the axial
flow. This can explain a drop by a factor of three in the laser
intensity (assuming a gain of $0.75cm^{-1}$, \cite{BenKish2001}).
The escape velocity to vacuum (i.e. through the exit hole in one
of the electrodes) is given by:
\begin{equation}
V_e=\frac{2}{\gamma-1}\times C_i\simeq7500m/s
\end{equation}
Hence, the gas jet that escapes the capillary during the pre-pulse
propagates about $3cm$ outside the capillary, and cause additional
absorption of the laser radiation. These two effects both cause
strong decrease in the laser pulse intensity, as observed in the
experiments. The axial flow of gas may also explain the increase
in the laser pulse duration, as it causes slightly different pinch
times along the plasma column, effectively broadening the laser
pulse.\\
In conclusion, we studied the effect of the pre-pulse current on
various aspects of capillary discharge X-ray lasers. As can be
expected, low amplitude pre-pulse current ($5A$) results in a non
uniform heating of the gas during the pre-pulse, initiating
instabilities during the magneto-hydrodynamic collapse. High
amplitude pre-pulse currents ($50A$) results in a uniform heating
of the gas and stable collapse of the plasma column. However, we
found that the pre-pulse also affects the properties of laser
pulse itself. Increasing the pre-pulse amplitude from $50A$ to
$100A$ mitigates the intensity of the laser pulse by an order of
magnitude. The pre-pulse also affects the duration of the laser
pulse - exhibiting longer pulses at higher pre-pulse currents. We
attribute both these effects to the axial flow of gas in the
capillary during the pre-pulse. Because the strong effect of the
pre-pulse on the laser properties it is important to take special
care in optimizing the pre-pulse in capillary discharge X-ray
lasers. \\This work was partially supported by the Fund for
Encouragement of Research in the Technion. We acknowledge the
technical assistance of Uri Avni, Shalom Aricha, and Yoav Erlich
in implementing the soft x-ray laser experiment.

\end{document}